\documentclass[Afour,sagev,times]{sagej}

\usepackage{moreverb,xurl}

\usepackage[colorlinks,bookmarksopen,bookmarksnumbered,citecolor=red,urlcolor=red]{hyperref}
\newcommand\BibTeX{{\rmfamily B\kern-.05em \textsc{i\kern-.025em b}\kern-.08em
T\kern-.1667em\lower.7ex\hbox{E}\kern-.125emX}}

\def\volumeyear{2026}

\usepackage{float}
\usepackage{subcaption}

\begin{document}

\runninghead{Africa‑Wide ARISS experiment}

\title{A First Africa‑Wide Amateur Radio Contact Experiment with the International Space Station}

\author{Domingos Barbosa\affilnum{1,3,6}, Miracle Chibuzor Marcel\affilnum{2}, Declan Kirrane\affilnum{5,6}, Miguel Avillez\affilnum{3,4}, Amare Abebe Gidelew\affilnum{7}, Carla Sharpe Mitchell\affilnum{8,9}, Vidvuds Beldavs\affilnum{10,11}, Claudio Ariotti\affilnum{12}, Stefan Dombrowski\affilnum{12}, Peter Kofler\affilnum{12}}

\affiliation{\affilnum{1}ESTGD, Instituto Politécnico de Portalegre, Campus Politécnico 10, 7300-555 Portalegre, Portugal\\
\affilnum{2}Pan-African Citizen Science e-Laboratory, 3004-516 Abuja, Nigeria\\
\affilnum{3}High Performance Computing Chair, University of Évora, R. Romão Ramalho 59, 7000 Évora, Portugal\\
\affilnum{4}Zentrum f\"ur Astronomie und Astrophysik, Technische Universit\"at Berlin, Hardenbergstrasse 36, 10623 Berlin, Germany\\
\affilnum{5}ISC Intelligence in Science, Rue du Trône 4, 1000 Brussels, Belgium\\
\affilnum{6}AERAP - Africa-Europe Science Collaboration and Innovation Platform, Rue de la Science 14,1040 Brussels, Belgium\\
\affilnum{7}African Astronomical Society, 1 Observatory Rd, c/o South Africa Astronomical Observatory, 7925 Cape Town, South Africa\\
\affilnum{8}SARAO - South African Radio Astronomy Observatory, Liesbeek House, River Park, Gloucester Road, Mowbray, 7700 Cape Town, South Africa\\
\affilnum{9}Foundation for Space Development Africa, Century Square, 8005 Cape Town, South Africa\\
\affilnum{10}University of Latvia, Riga Photonics Centre, \v{S}\c{k}\=u\c{n}u iela 4, LV1050 Riga, Latvia\\
\affilnum{11}ACES Worldwide - Alliance for Collaboration in the Exploration of Space, Virginia, USA\\
\affilnum{12}ARISS - Amateur Radio on the International Space Station, EU\\
}

\corrauth{Domingos Barbosa, ESTGD, Instituto Politécnico de Portalegre; HPC Chair, University of Évora, Portugal}

\email{domingos.barbosa@uevora.pt}

\begin{abstract}
The Amateur Radio on the International Space Station (ARISS) program is recognized by its great Science, Technology, Engineering, and Mathematics (STEM) impact: it brings space science to education and outreach initiatives by enabling direct contact with astronauts in orbit and other space savvy professionals. Strategic partnerships between ARISS and digital platforms for citizen science act as a powerful real-world STEM motivator and can overcome logistical barriers, demonstrating a scalable model for inclusive outreach, sparking curiosity in space exploration and STEM careers among underrepresented communities. 

As a case study, this article focus on the impact of a historic milestone: on April 18$^{th}$ 2025 the Pan-African Citizen Science e-Laboratory (PACS e-Lab) orchestrated the first Africa-wide ARISS contact experiment and connected 300 students, educators, and STEM enthusiasts from 38 African nations and beyond with NASA astronaut Nichole Ayers aboard the ISS Expedition 72. Within the African context, this initiative not only inspired a new generation of African scientists, but it also set a precedent for wider collaboration on space science education across the continent. 

Strategic partnership between ARISS and PACS e-Lab promotes capacity building through international and regional cooperation on space science and adds a new "New Space" narrative that emphasizes the interconnectedness of space with our daily lives and the possibility that everyone has to be a part of it. 
\end{abstract}


\maketitle

\section{Introduction}

The International Space Station (ISS) is rightly  considered  one of humanity’s great cooperative and engineering achievements. Built and operated by 20 countries over nearly 30 years, the ISS has been the orbital home to more than 290 crew members from 29 countries and has been occupied continuously since November 2000\endnote{\url{https://www.nasa.gov/reference/international-space-station/?utm_source}}. Its international crews and their activities, occupancy and equipment are governed by treaties between participant nations. The ISS mission is to conduct research and development in low Earth orbit (LEO) to learn how humanity can better live and work in space and to return the benefits of this research to people on the ground. At the end of the space station’s useful life in 2030, NASA and its international partners will deorbit the station, safely disposing the space station into the Nemo point, a remote  part of the Pacific Ocean at 2,688 km away of the nearest land \cite{NASA2024ISSDeorbit}. 

ISS socioeconomic impact is immense  and it is observed that ISS-based studies, particularly those involving public-private partnerships, register higher scientific impact than similar Earth-based research studies \cite{npjMicrogravity2025,ESA1}. But its success goes well beyond economics: the ISS strengthens international relationships and inspires future generations, detailed in reports covering 30 years of data \cite{MAYOROVA2014147}. ISS is rightly recognized as an innovative educational platform for space research and aerospace education helping to address the international challenge of providing qualified aerospace workforce and support educational programs at training specialists capable of solving tasks related to the development, creation, testing and utilization of complicated technical systems \cite{MAYOROVA2014147}. For instance, 34\% of educational kits made by ESA for primary and secondary schools are inspired by Astronauts and ISS missions \cite{ESA2} and several national educational programs included ISS in their STEM activities. For example, "The ISS: {\bf I}nnovatio, {\bf S}cientia, {\bf S}apientia"\endnote{The project’s acronym (ISS) also stands for "Innovation, Science, Wisdom" in Latin} was a national STEM/space education competition in Italy (2015–2018) project, promoted by the Ministry of Education, Ministry of Defense, and Italian Space Agency targeted at secondary school students to develop innovative experiments for the International Space Station (ISS)\cite{Bonacci_2023}.

The ISS cooperation policy includes the Amateur Radio on the International Space Station (ARISS) experiment, a key cooperation experiment for education and outreach that enables students worldwide to engage directly with astronauts through a radio scientific experiment via ham radio. With time passing, ARISS experiment program became a powerful, "soft power" instrument of space cooperation policy by fostering international collaboration, promoting interests in STEM education, and strengthening public engagement with the ISS program through unique, real-time interactions with astronauts aboard the ISS \cite{Bauer2019,Diggens2023,Conley}. Through partnership between international space agencies (NASA, ESA, CSA, JAXA, Roscosmos) and amateur radio organizations, ARISS operates continuously to connect crew members with students and communities worldwide, promoting peaceful, cooperative use of outer space. The goals of the ARISS program are\endnote{from \url{https://www.ariss.org/}}:
\begin{itemize}
{\it 
    \item Inspire an interest in science,technology, engineering, and math(STEM) subjects and in STEM careers among young people.
    \item Provide an educational opportunity for students, teachers, and the general public to learn about space exploration, space technologies, and satellite communications.
    \item Provide an educational opportunity for students, teachers, and the general public to learn about wireless technology and radio science through Amateur Radio.
    \item Provide a contingency communications system for NASA and the ISS crew.
    \item Provide crew with another means to directly interact with a larger community outside the ISS,including friends and family.
    }
\end{itemize}

Since its inception in November 2000, ARISS has maintained continuous operations, hosting $\sim 60-100$ contacts, involving $\sim$15,000 to 100,000 students. Indeed, the ARISS opportunities have proven themselves as a tremendous educational boom. The events are accomplished using the ARISS amateur radio systems on ISS, through the support of ISS crew members that have obtained their amateur radio licenses and through hundreds of ARISS international volunteers around the world. These volunteers mentor the schools, help set up ham radio equipment in the schools, and then prepare the students to conduct the contact with the ISS crew.

\subsection{Space education for the African New Space era}

The development of Astronomy and Space Science (A\&SS)in Africa has grown significantly over the past few years and the African landscape has changed dramatically following the African Union (AU) steps to promote the development of A\&SS on a continental scale for improving some of the main socioeconomic and environmental challenges that Africa is facing, and for achieving United Nations Sustainable Development Goals (UN SDGs) \cite{2018NatAs...2..507P}. These steps significantly expanded the International Astronomical Union (IAU) and the Organisation for Economic Co-operation and Development (OECD) recommendations for inclusion and development of cross-curricular themes for high-school curricula, namely those related to environmental education, global citizenship, career education to include space topics \cite{retre2019big,organisation2020curriculum}. A regional milestone, the African Astronomical Society (AfAS) launched itself as a focus point and a new professional society to bridge teaching and research. Hence, AfAS contributes to Human capital Development and its follows the path set by AU strategies that includes the advancement of the African Space Agency \cite{Leeuw_Govender_Takalana_Randriamanakoto_Mamo_2019,Tessema2023AfAS}.

Africa became home to landmark astronomy projects like the global Square Kilometre Array (SKA) MID telescope \cite{2017ARep...61..288G,wild2017giant} and its precursor MeerKAT \cite{2012AfrSk..16..101B}, installed in South Africa. These projects aggregate the SKA African Partner countries, ie Botswana, Ghana, Kenya, Madagascar, Mauritius, Mozambique, Namibia, and Zambia. These nations collaborate with South Africa to support the expansion of the SKA Observatory and often work through the African Very Long Baseline Interferometry Network (AVN) program \cite{2014arXiv1405.7214G}, aiming to develop radio astronomy infrastructure and capabilities across the continent. The AVN significantly improves the science capabilities of the global VLBI community \cite{Venturi2021VLBI} and it contributes to the regional space infrastructure supporting space navigation, geodesy and monitoring of the near-Earth space. Besides these large scale projects, optical observatories have recently spread in the African landscape like the TRAPPIST-North in Morocco Atlas mountain \cite{Barkaoui2017} fit for exoplanet research and high energy observatories like The High Energy Stereoscopic System (HESS) \cite{P_hlhofer_2023} in Namibia, a system of imaging atmospheric Cherenkov telescopes for the investigation of cosmic gamma rays and already operating since 2002. 

Indeed, these developments led to important capacity building and outreach efforts in astronomy and space science, namely the Development in Africa with Radio Astronomy (DARA) program that offers training programs and postgraduate scholarships for African students in STEM, astrophysics, big data and space that targets primarily at citizens of AVN partner countries (e.g., Botswana, Ghana, Kenya, Madagascar, Mauritius, Mozambique, Namibia, Zambia) \cite{2018NatAs...2..507P,2018NatAs...2..505H}. These efforts were complemented by other regional projects like the Development of PALOP Knowledge in Radioastronomy project (DOPPLER), an international capacity-building initiative focused in Mozambique \cite{Ribeiro_Paulo_Besteiro_Geraldes_Maphossa_Nhanonbe_Uaissone_2009,10.1007/978-3-030-67411-3_10}. DARA and DOPPLER programs paved the way for careers in space and technology in Africa. Indeed, the nature of the global space agenda requires a sustained capacity building to build knowledge and skills in space technology and policy in particular within underrepresented communities.

Recently, ARISS and the Pan-African Citizen Science e-Laboratory (PACS e-Lab) partnered to host a first communication experiment on April 18$^{th}$ 2025 with ISS astronauts, at a momentous week for African space science. The event happened just after the commemorations of the International Day of Human Space Flight (12th April 2025), remembering Yuri Gagarin's historic 1961 spaceflight — the first human journey beyond Earth's atmosphere - and during the week of the official inauguration of the African Space Agency, a major milestone for space affairs in the African continent. Through these activities, partnership between PACS e-Lab and ARISS enhanced opportunities for the school community (students, teachers, families and community members) to become more aware of the substantial benefits of human spaceflight and the exploration and discovery that occur on spaceflight journey through the exploration of amateur radio \cite{ARISS2025,Diggens2023,Kramer2002}. 

As described extensively in \cite{WEIZMAN2025101708}, New Space technological innovations led to recommend new pedagogical approaches and skills fostering new STEM contexts \cite{retre2019big, freeman2021stem,rosu2023space}, integrating online and international education. This also requires adding economical aspects as a key ingredient for space education in the 21st century \cite{pelton2004needs}. These advancements make the United Nations Sustainable Development Goals more achievable, and open up the possibility of new beneficial collaborations. As noted in \cite{WEIZMAN2025101708}, the "Astronomy Education Paradigm" was good for the "Old Space" time, but the "New Space" era requires rethinking the needs and goals of the field of space education \cite{2018NatAs...2..507P,AMBROSIUS2023101535}. Naturally, these PACS e-Lab and ARISS activities with direct contact with astronauts through a space communication setup do fit into the broader picture of STEM activities that benefit the New Space paradigm by bringing knowledge and tools to underrepresented communities. 

We acknowledge most discussions of space education focus on STEM fields to provide the knowledge capacity to analyze space travel, cargo and people transport to orbit and the support to human life under space extreme conditions. But advanced education on space exploration topics should address as well social science perspectives including public policy development, economics, sociology, psychology, and anthropology to better acclimate our national and international policies, societies, and economies in particular those from underrepresented regions to the emergent mass-scale orbital occupancy. And that is why ISS and ARISS wide scale experiments are such a valuable instruments to enthuse newer generations to the new Space era.

\section{PACS e-Lab and African ARISS initiatives}\label{PACS e-Lab and African ARISS initiatives}

PACS e-Lab is a non-profit organization/platform dedicated to advancing STEM education, promoting citizen science, and enabling entry-level research participation through hands-on activities in astronomy and space science across Africa. The lab utilizes both internal and external collaborations, digital technologies, e-learning/e-science, volunteers, and innovative projects to democratize astronomy and space science across the continent. PACS e-Lab embraced the new astronomy paradigm for STEM that is shifting from a classical observational science to a dynamic, interdisciplinary approach. In the new space era, "Space Education" is redefined by integrating astrophysics with computational thinking, data literacy, and engineering, using the cosmos as a sandbox to solve complex terrestrial problems \cite{AMBROSIUS2023101535}. Hence, the platform has conducted successful science campaigns in about 50 countries in Africa and beyond, engaging thousands of teachers, students, and STEM enthusiasts in citizen science, research, and space-based STEM engagement activities\endnote{\url{https:\\www.pacselab.space}}. 
We highlight below some of the PACS e-Lab collaborations with global organizations on citizen science with an astronomy and space exploration focus:
\begin{itemize}
    \item {\bf Asteroid Searches}: Partnering with the International Astronomical Search Collaboration (IASC), volunteers analyze telescope data from the Panoramic Survey Telescope and Rapid Response System (Pan-STARRS) \cite{2016arXiv161205560C} and the Catalina Sky Survey \cite{larson1998catalina,FARNOCCHIA2016327}, a NASA funded project supported by the Near Earth Object Observation Program (NEOO) under the Planetary Defense Coordination Office (PDCO)\endnote{\url{ https://science.nasa.gov/planetary-defense/}}. These survey programs are used to detect new asteroids. Participants are trained to submit their astrometric reports to the Harvard coordinated IAU Minor Planet Center\endnote{\url{https://iasc.cosmosearch.org}} \cite{miller2012,2025LPICo3088.5003M}.
    
    \item {\bf Exoplanet Observations}: major collaborator to the NASA Exoplanet Watch, steering African Citizen scientists to observe known exoplanets registered in space mission databases and help refine their transit times \cite{2025JAVSO..53..135M}.This supports major follow-up science from several observatories including the James Webb Space Telescope (JWST)\cite{Zellem_2020}.
    
    \item {\bf SkyMapper}: main African collaborator platform to the SkyMapper program, which aims to build a connected, decentralized global network of telescopes that can be operated remotely for scientific research, education, outreach, and commercial observations\cite{2025epsc.conf.1038M,2026AAS...24713704E}
    
    \item {\bf Astrophotography \& Deep Space Image Processing}: Volunteers process raw, open-source data from major observatories like Las Cumbres Observatory (LCO) Global Telescope Network \cite{2013PASP..125.1031B}, the Hubble Space Telescope, and the JWST to create professional-grade visual imagery and acquire know-how and hands-on training on advanced digital processing techniques\cite{2006SSRv..123..485G,Scoville_2007}.

    \item {\bf Double-Star Research}: Participants use historical records and major astronomical databases from flagship space missions like the ESA mission Gaia \cite{2016A&A...595A...1G} to measure and follow-up binary star systems and other variable star systems \cite{2024JDSO...20...39M}.

    \item {\bf Observational astronomy \& Citizen Science}: engagement of learners accross Africa in the use of robotic telescopes like the Las Cumbres Observatory\cite{Brown_2013}, Slooh\cite{gershun2014mixed}, and MicroObservatory\cite{sadler2001,gould2006educational}.

    \item {\bf Telescope Distribution}: support for local STEM groups to acquire handy telescopes for outreach and teacher training as an entry gateway to astronomy \& space science\endnote{\url{https://www.ssvi.be/}}.

    \item {\bf Astronomy lectures}: collaboration with Free AstroScience\endnote{\url{www.freeastroscience.com}} to deliver public and free astronomy and space science concepts via lessons on social media and podcasts. 

\end{itemize}

Recently, the group has adopted ARISS as a platform for continued public engagement and expansion to space science education \cite{2025afas.confE..57M, 2025LPICo3088.5003M}. Indeed, as in \cite{MAYOROVA2014147,WEIZMAN2025101708}, ARISS experiments promote transitioning an interdisciplinary approach that integrates space-related topics with digital and technological literacy.

Historically, individual ARISS events had been organized in South Africa
\endnote{\url{ https://www.africaninspace.com/education/ham\_radio.shtml}}$^,$\endnote{\url{https://amsat-uk.org/about/history/first-african-in-space/}} and Morocco
\endnote{\url{https://community.libre.space/t/ariss-contact-higher-national-school-of-computer-science-and-systems-analysis-ensias/1565}}
for decades, but with limited Pan-African coordination. PACS e-Lab leveraged from its past successes and vast network and advanced to support ARISS experiments to promote its hands-on adoption across institutions throughout Africa. 

The benefits of these PACS e-Lab initiatives enable more underrepresented communities within its reach to access this high-impact space science experiments, fosters interest in space exploration, and provides a platform for cross-cultural, Pan-African scientific engagement. They have a recognized social and economic impact that contribute to the development as outlined through the United Nations’ Sustainable Development Goals (SDGs). The SDGs impacted by these related science education activities involving the support of current powerful digital technologies include at least SDGs 4, 5, 8, 9 and 10: quality education; gender equality; decent work and economic growth; industry, innovation, and infrastructure; reduced inequalities. This is paramount to successfully build citizen science networks that foster STEM enthusiasm across underrepresented activities \cite{BARBOSA2022612}.

\subsection{The ARISS communication experiment setup}

ARISS operates through five regional offices : ARISS-US, ARISS-Canada, ARISS-Japan, ARISS-Russia, and ARISS-Europe, coordinating efforts of about fifteen countries. Submission windows typically open from February to March and September to October each year \cite{Conley} and local radio groups orchestrate activities locally with their respective space agency, e.g. NASA, CSA, JAXA, Roscosmos and ESA.

This program builds on the legacy of more than two-dozen space shuttle missions since 1983 to enable space communications with virtually amateur radio stations all around the world. This ARISS predecessor program called SAREX (The Space Shuttle Amateur Radio Experiment) enabled dozens of astronauts in orbit to address thousands of kids in school and to their families on Earth while they were in orbit \cite{Conley}. 

We note the ISS decommission by 2030 and NASA's announced transition to support the development of commercial, private space stations to continue research in Earth's orbit represents a major pivot point for STEM activities including outreach: it will likely end student-designed experiments and STEM partnerships with the ISS National Laboratory and future STEM research or activities will likely be commercial, driven by private-sector interests. Yet, ARISS legacy constitute a powerful channel to enthuse younger generations about space science that must be kept and evolved within the New Space paradigm. 

After the ISS decommissioning, Tiangong Chinese space station will become the longest continuously inhabited space station in operation since its first crews by 2022. The CSSARC, the Amateur Radio payload for the Chinese Space Station, proposed by the Chinese Radio Amateurs Club (CRAC) can, hopefully, build and expand on the ARISS legacy for the benefit of international cooperation and STEM outreach.

ARISS interactions unfold in the form of a live question and answer contact with an astronaut or cosmonaut aboard the International Space Station (ISS) following the principles outlined through integration of the ARISS School Program into the school curricula as an external intervention. Opportunities are provided and worked out for school students or youngsters to ask their own prepared questions of the astronauts using a two-way Amateur Radio, in a {\it No Roadblocks in Low Earth Orbit} mode \cite{MAYOROVA2014147,Diggens2023No}.

The ARISS Operations Team is made up of ARISS Technical Mentors, scheduling/technical representatives, and an orbital prediction specialist.  These ARISS Technical Mentors are the experienced radio amateur volunteers who work with the schools, teachers, and local Amateur Radio groups that actually make the contacts with the ISS astronaut team. Scheduling/technical representatives work within the space agencies, primarily NASA in the USA and Roscosmos in Russia, to secure the final schedules for the contacts and the astronaut training setting up final preparations well in advance.


The ARISS team has developed various hardware elements for the ISS amateur radio station \cite{Conley}. These hardware elements have flown to ISS on three Shuttle flights and one Progress flight. The initial educational outreach system consists of an FM radio system attached to some externally mounted antennas operating in VHF (144-146 MHz) and UHF (435-438MHz) frequency bands. This initial radio system is located in the Functional Cargo Block, named Zarya. This system supports FM voice operations and packet radio (computer-to-computer radio-link) capabilities. Packet radio has several capabilities including an Instant Messaging type system and a Bulletin Board System that allows radio amateurs to store and forward messages and allows the orbiting crew to send e-mail to all hams or to individuals.

To improve communications, ARISS has developed the Interoperable Radio System (IORS) \cite{IORS}, the ARISS next-generation radio system on ISS lown to ISS on SpaceX CRS-20 and installed in the ISS Columbus module by Expedition 63 Commander, Chris Cassidy on September 2, 2020 and Direct Contact Ground Station Requirements and Worldwide Telebridge Facilities that facilitate the ISS communication. 

The Worldwide telebridge system consists of an international network of ISS ground stations that can be linked to school groups using nowadays a digital over the internet conferencing system stemming from Direct Contact Ground Station Requirements. The ISS is a near-polar orbiting object with an orbital inclination set at 51.6 degrees measured at the equator and an ARISS ground station is essentially a directional antenna system, capable of azimuth and elevation motion, accurately steered using computer-controlled hardware. This enables the required precise tracking of the ISS enabling communication quality with good Signal to Noise ratio during the ARISS event until ISS Loss of Signal (LoS).

\section{An Out of Africa case study: PACS e-Lab - ARISS experiment}

For the agreed ARISS experiment, PACS e-Lab’s tapped its grassroots connections with several underrepresented STEM communities and individuals across the continent, after their previous projects like asteroid hunting \cite{2025afas.confE..57M}, a key component of planetary defense efforts. These projects engaged thousands of citizen scientists across more than 50 African countries and some of these citizen scientists have successfully detected and reported hundreds of asteroids, reported to the Minor Planet Center\endnote{\url{https://www.minorplanetcenter.net/iau/mpc.html}}, where astronomers worldwide collaborate to track their orbits and calculate any potential collision probabilities.

PACS e-Lab event proposal was previously selected through the ARISS-Europe application cycle for the January to June 2025 window. Due to the fast ISS orbital speed (an average of $\sim4^\circ$/minute), the communication window with ISS spans typically for 10-12 minutes over a ground station and everything has to be timed and well tested to enable future ARISS events. The standard participation for interactions with astronauts and ground station team is set to 20-30 individuals/questions selected through mandatory vetting procedures by the ARISS team and NASA. In the selection process, the selection committee prioritized the uniqueness of questions, gender balance, and representation from diverse countries in Africa and the Middle East (with two participants from Afghanistan and the United Arab Emirates). 

Recognizing that this was a once-in-a-lifetime opportunity, PACS e-Lab ensured that as many young people as possible were included, giving them the chance to be inspired through active participation as members of the wider audience. Additionally, to build excitement and deepen participants’ understanding of the ISS and space science activities, PACS e-Lab shared a series of space facts through social networks including instructions to use popular ISS tracking software\endnote{\url{https://spotthestation.nasa.gov/}}$^,$\endnote{\url{https://www.heavens-above.com/}}. The Official schedule and selected ARISS ground station for the telebridge contact with NASA astronaut: 

\begin{itemize}
    \item Option \#21: Friday, 2025-04-18, 10:26:16 UTC, 76º elevation via ground station IK1SLD;
    \item ISS astronaut scheduled for contact:  Nichole Ayers (KJ5GWI), NASA;
    \item The operator of the ARISS ground station: IK1SLD - Claudio Ariotti; 
    \item ARISS moderator for the telebridge contact: ON6TI - Stefan Dombrowski.
\end{itemize}

\begin{figure}[ht!]
\centerline{
 \includegraphics[width=0.5 \textwidth]{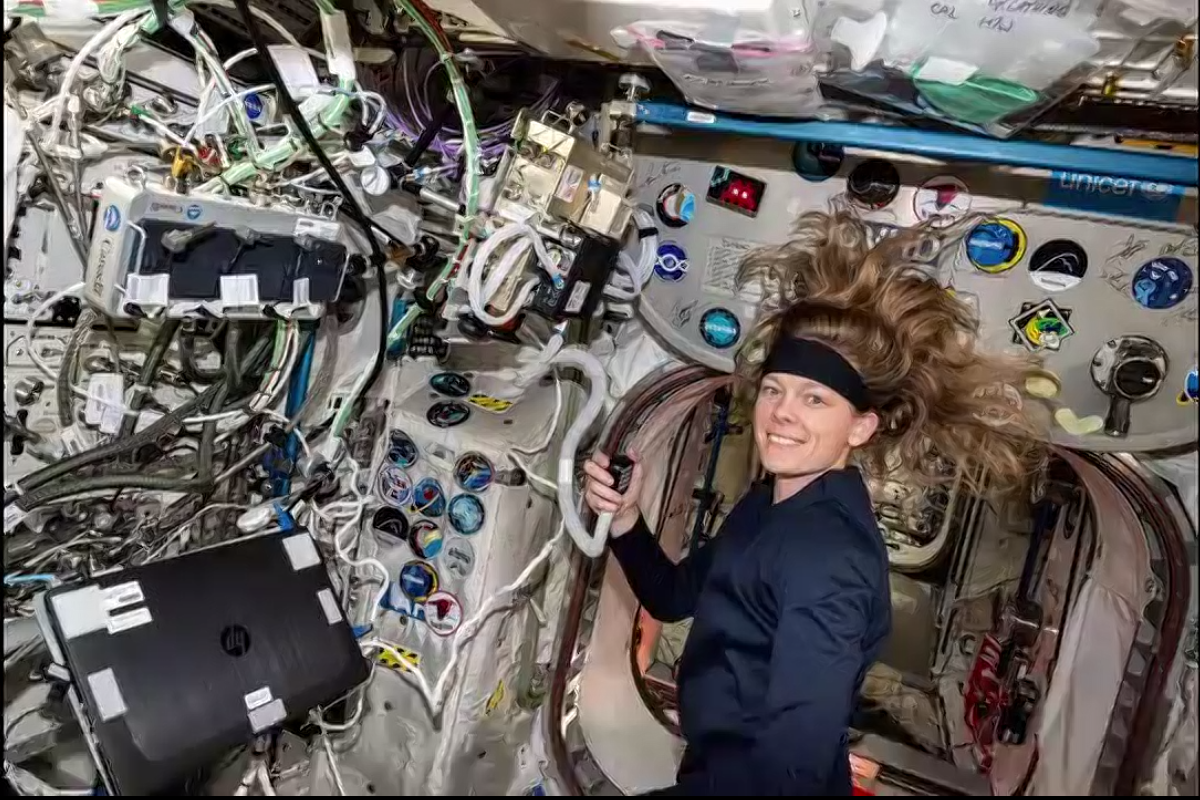}}
    \caption{NASA Astronaut Expedition 72 Flight Engineer Nicole Ayers aboard ISS; talking to students from a USA Arkansas school a few days before the ARISS wide Africa experiment. Credit: NASA Johnson Space Center, flickr iss072e882080 (March 31, 2025).}
    \label{fig:ayers}
\end{figure}

\begin{figure}[ht!]
\centerline{
  \includegraphics[width=0.5 \textwidth]{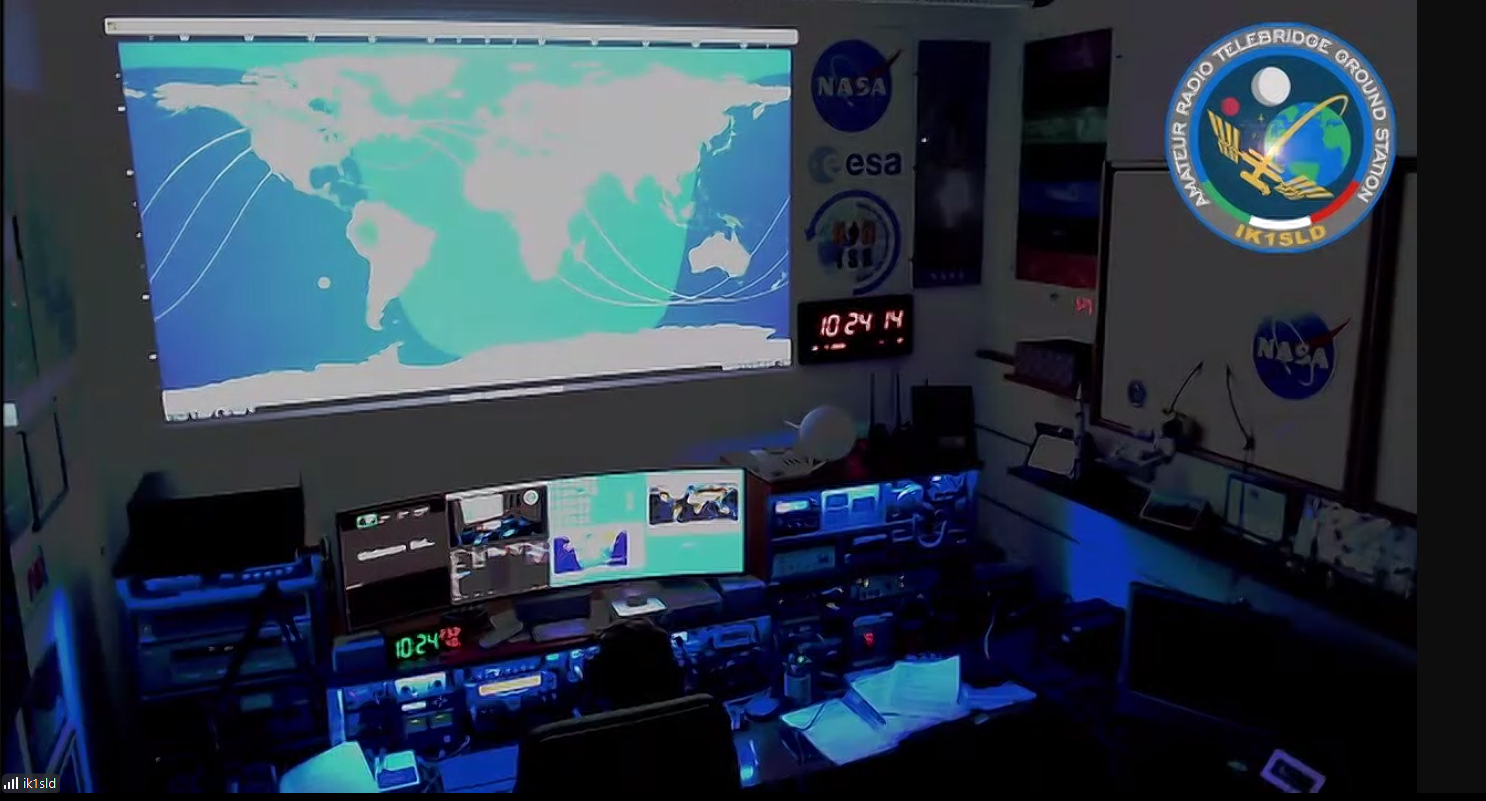}}
    \caption{ARISS Telebridge broadcasting station operated by K1SLD - Claudio Ariotti, Casale Monferrato (Italy). The ground station broadcasted ISS communications via internet to participants during the event until ISS Loss of Signal. Credit: K1SLD - Claudio Ariotti, ARISS, photo obtained during the event.}
    \label{fig:telebridge}
\end{figure}


\section{Results}

\subsection{Event Execution}

The scheduled ARISS contact successfully took place within the assigned space radio window of April 18, 2025. As expected, the contact lasted approximately 10 minutes once radio contact with the ISS was established, during which students and young enthusiasts from across Africa and the Middle East regions engaged in a live real-time Q\&A session with NASA astronaut Nichole Ayers reading their own prepared questions and listening to the answers from the astronaut before ISS Loss of Signal (LOS), at which time the ISS has traveled beyond the visible horizon of the receiving ground station. The session was facilitated via the ARISS ground station IK1SLD and retransmitted via internet. ARISS ground station clear signal reception was confirmed by many other ground stations from Ireland to central Europe that also received and listened ISS communications.

Very high ARISS excitement levels were exhibited by the students across African nations participating in the experiment, creating a vibrant and positive atmosphere about space sciences. There were almost ten times as many students, young participants and citizen scientists assisting than the selected 31 participants for astronaut-ARISS public communication. Senior participants and teachers unanimously reported that astronauts, male and female, are a significant positive role model for students and young citizen scientists, confirming the findings from \cite{Bonacci_2023,Diggens2023No,MAYOROVA2014147,AMBROSIUS2023101535}.

The questions covered a wide range of topics, including aspects of astronauts daily life (how astronauts drink water, sleep, maintain hygiene, and stay mentally healthy), scientific and technical subjects like how microgravity affects problem-solving, how astronauts withstand radiation, and whether microorganisms can survive on the ISS exterior. Broader themes were also explored, including planetary exploration, astrobiology, the fate of the ISS after decommissioning, and humanity’s potential to travel across the stars. Personal insights were shared too, with students asking about the astronaut’s inspiration, experiences with the view Effect, and emotional resilience in space, emphasizing that Social, political, and economic issues are as critical as technological ones for a space faring society \cite{GENTA2006287,AMBROSIUS2023101535}.

The session provided an intense, fast and engaging and inspiring glimpse into both the human and scientific aspects of life in orbit. The astronaut provided thoughtful, engaging responses that captivated both the students and the wider audience. This was followed by additional outreach and policy talks surrounding the event, all in live stream, recorded and uploaded to YouTube, allowing those unable to attend live to watch at their convenience\endnote{\url{ https://www.youtube.com/watch?v=dGViDNEA754}}.

\subsection{Reach and Impact}

\begin{figure}[ht!]
\center{
  \includegraphics[width = 0.5 \textwidth]{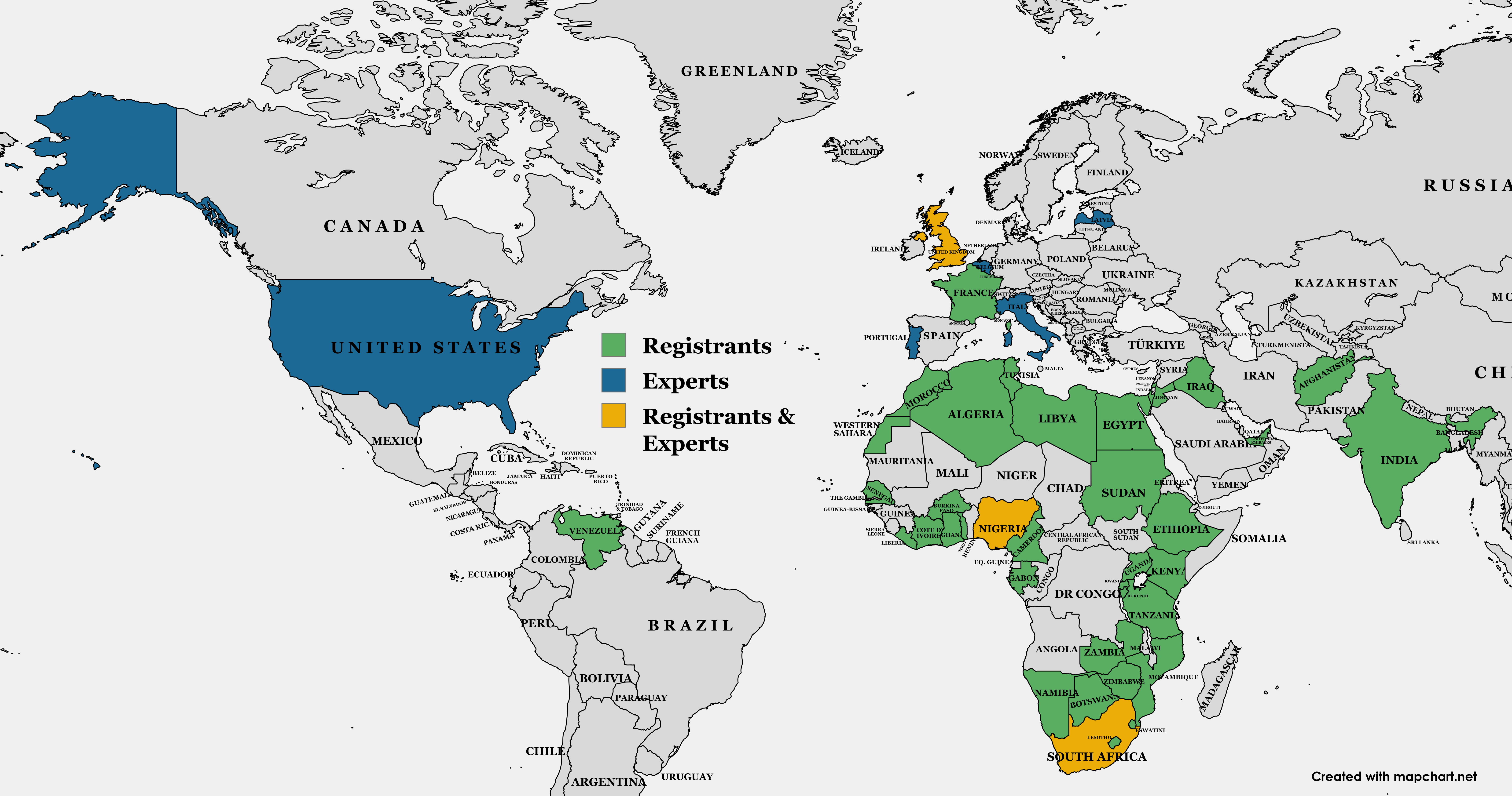}}
    \caption{Geographical distribution of registrants and Experts.}
    \label{fig:worldmap}
\end{figure}

\begin{figure}[ht!]
\centerline{
  \includegraphics[width = 0.5 \textwidth]{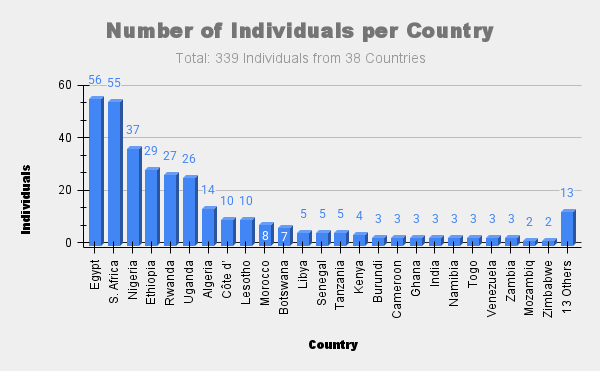}}
    \caption{Number of individual registrants per country.}
    \label{fig:number_individuals}
\end{figure}

\begin{figure}[ht!]
\centerline{
  \includegraphics[width = 0.5 \textwidth]{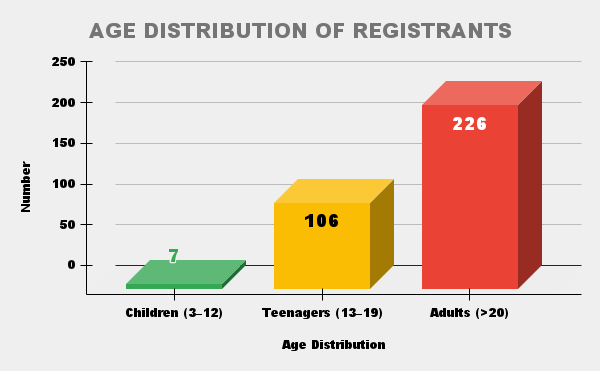}}
    \caption{Age Distribution of Registration.}
    \label{fig:age}
\end{figure}

\begin{figure}[ht!]
\centerline{
  \includegraphics[width = 0.5 \textwidth]{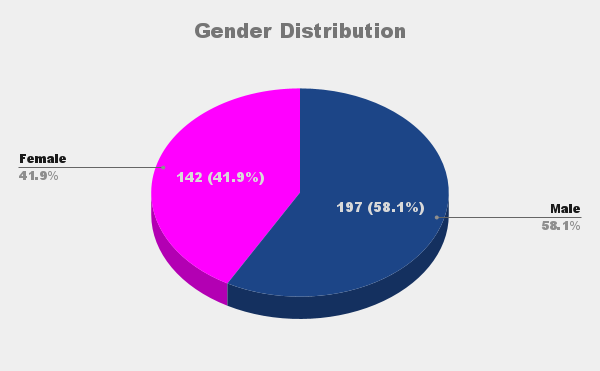}}
    \caption{Gender Distribution of registrations.}
    \label{fig:gender}
\end{figure}

This Pan-African ARISS event generated significant regional and international engagement. The event itself generated substantial traction on social media platforms. Campaigns on education and outreach channels in the most popular social networks like LinkedIn, Twitter (X), Facebook, Instagram, and YouTube collectively recorded several tens of thousands views or impressions, as testimony of impact and the great interest in space activities generated across the African continent and Middle East. Key outcomes include:
\begin{itemize}
    \item Geographical Reach: PACS e-Lab received registrations from about 339 individuals from 38 countries in Africa, Asia, Europe, and the Americas. These included, among others, Egypt, Nigeria, South Africa, Ethiopia, Rwanda, Uganda, Algeria, Lesotho, Côte d’Ivoire, Botswana, Morocco, Tanzania, Libya, Senegal, Namibia, Zambia, Cameroon, Ghana, Kenya, India, Burundi, Togo, United Arab Emirates, Zimbabwe, Mozambique, Venezuela, Liberia, Jordan, Afghanistan, Eswatini, Gabon, Iraq, Sudan, Tunisia, Burkina Faso, Israel, United Kingdom, and France, as in Figure\ref{fig:worldmap} and Figure\ref{fig:number_individuals};
    \item Geographical origin of selected astronaut-public Q\&A: Afghanistan, Algeria, Cameroon, Egypt, Eswatini, Ethiopia, Ghana, Lesotho, Liberia, Libya, Morocco, Namibia, Nigeria, Senegal, South Africa, Tanzania, Uganda, United Arab Emirates;
    \item Wide age distribution: from children to Adults, see histogram in Figure\ref{fig:age};
    \item Balanced gender distribution among registrants, supporting PACS e-Lab’s commitment to inclusive STEM activities, see Figure\ref{fig:gender}.
\end{itemize}

\begin{figure}[ht!]


\centerline{
\includegraphics[width = 0.2 \textwidth]{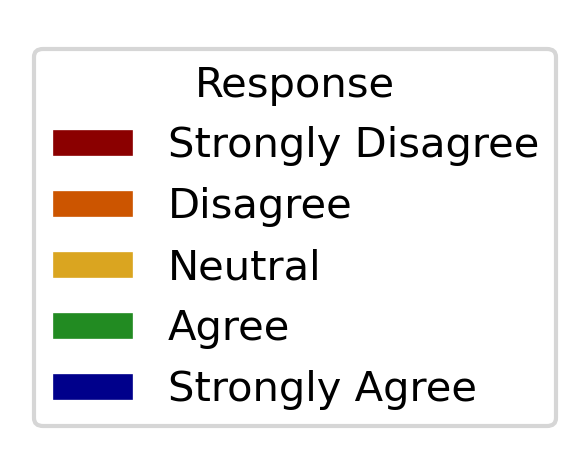}}

\includegraphics[width=0.5 \textwidth]{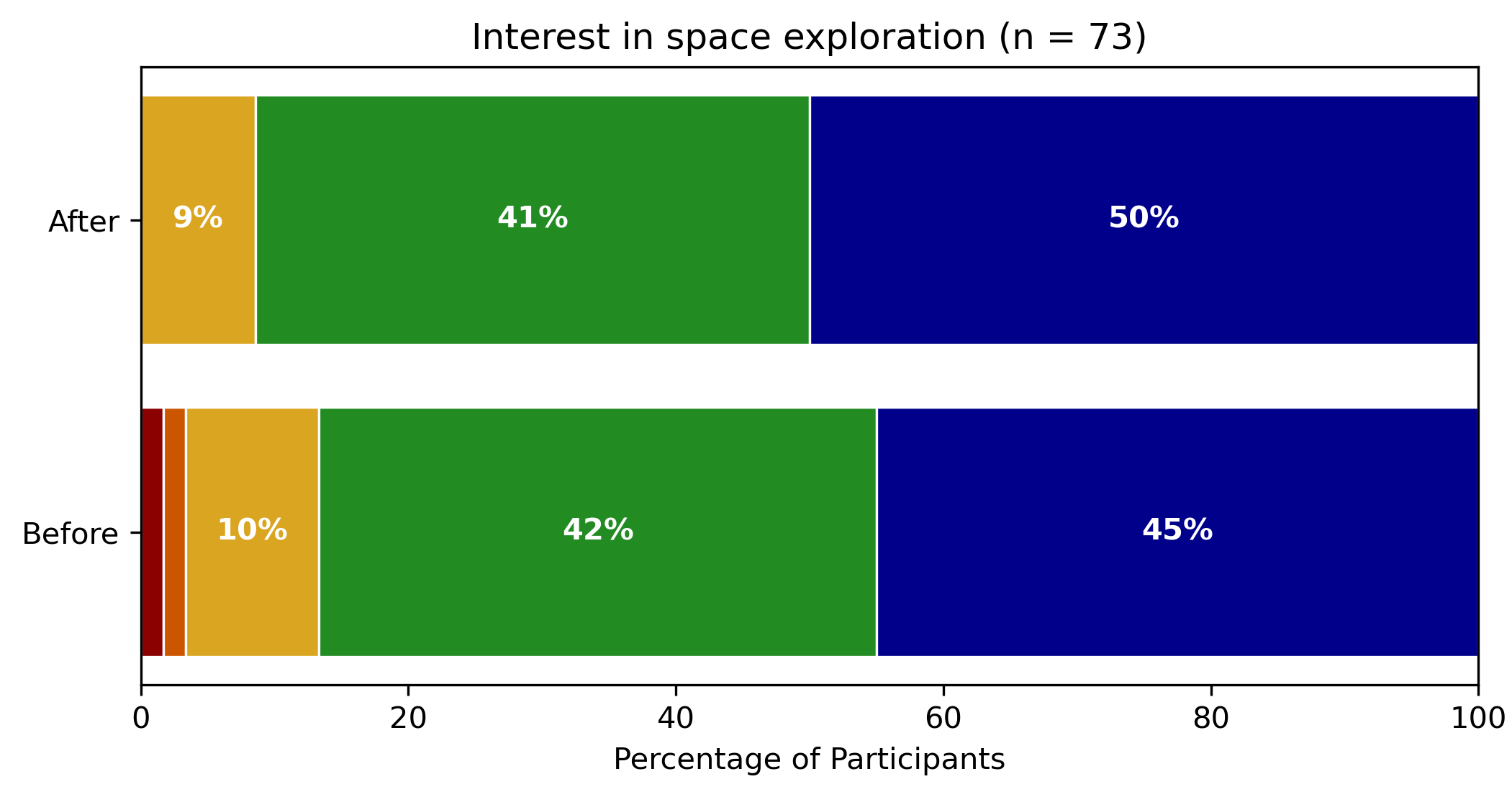} 

\includegraphics[width=0.5 \textwidth]{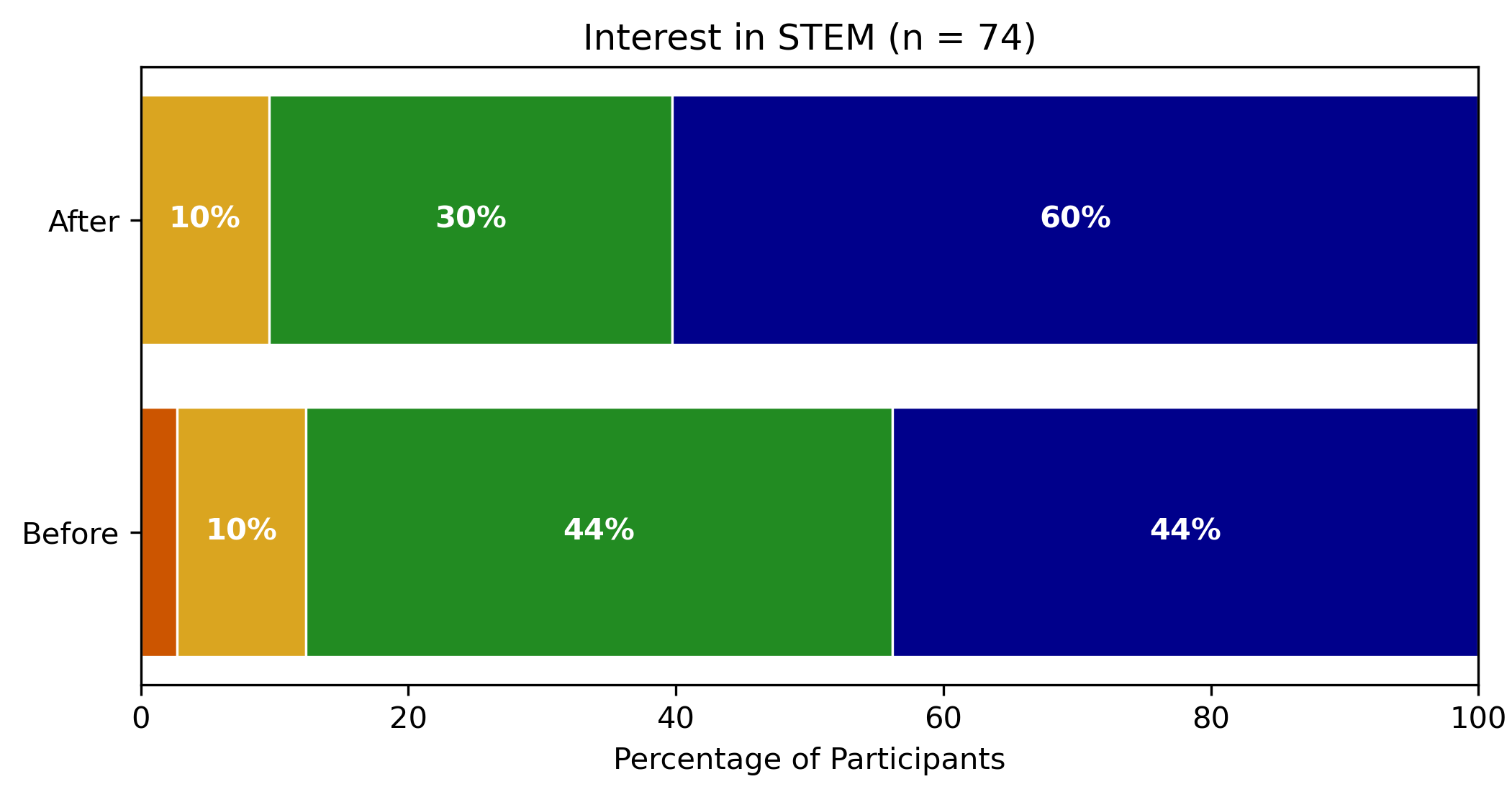} 

\includegraphics[width=0.5 \textwidth]{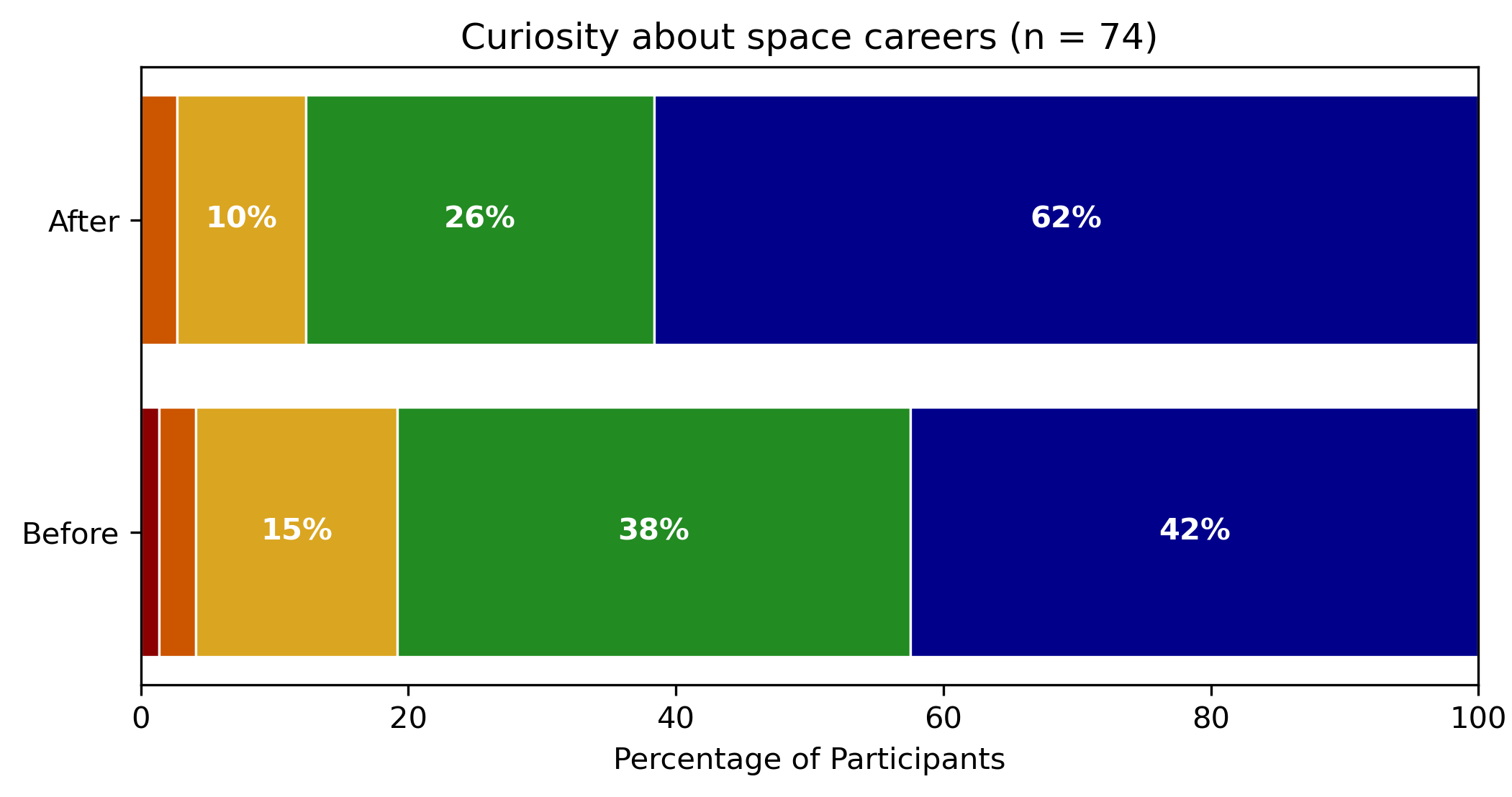}
\caption{Survey: Interest and impact on career prospects and Space activities.}\label{fig:graph1}
\end{figure}

\begin{figure}[ht!]
\includegraphics[width=0.5 \textwidth]{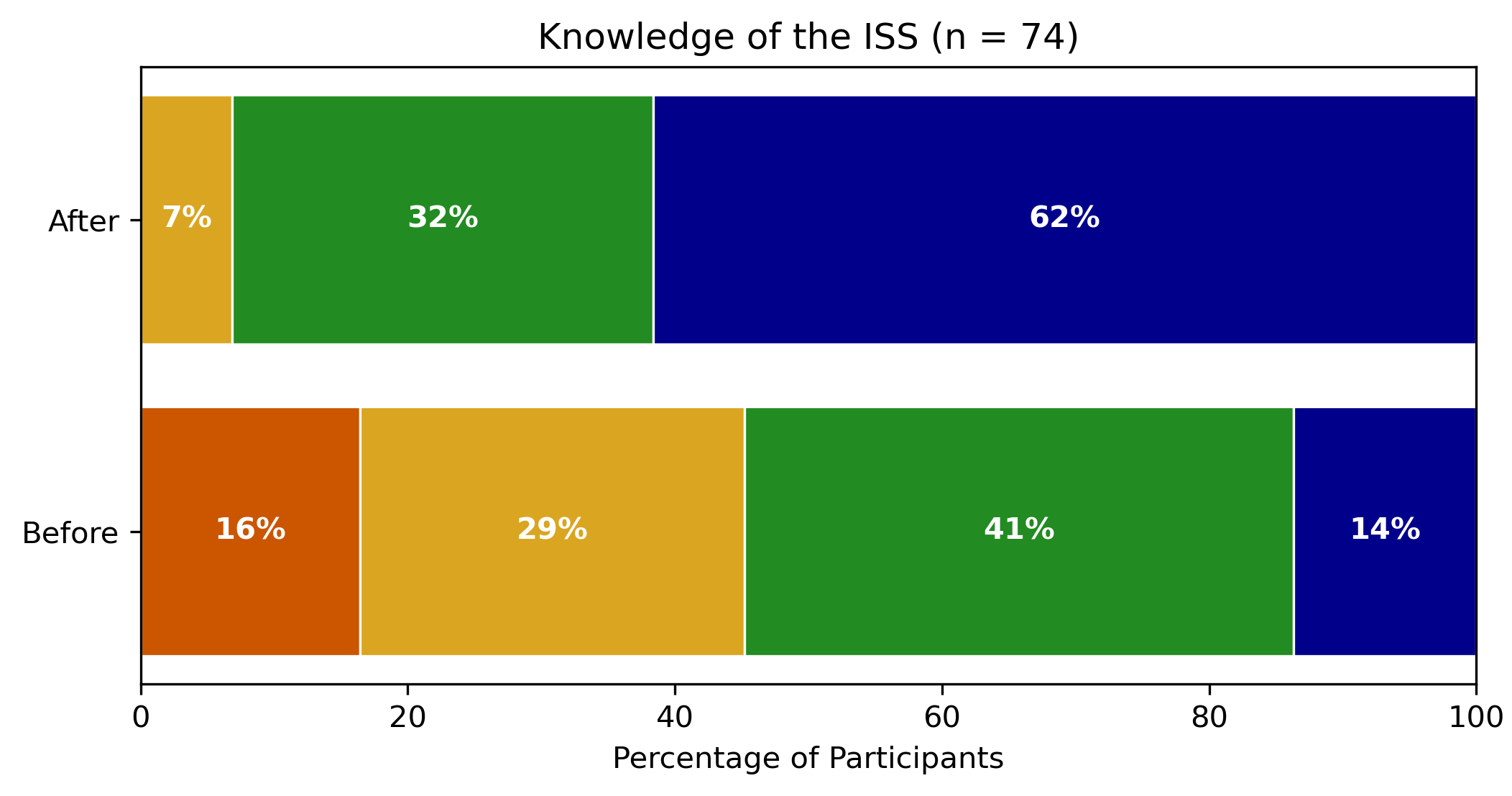}

\includegraphics[width=0.5 \textwidth]{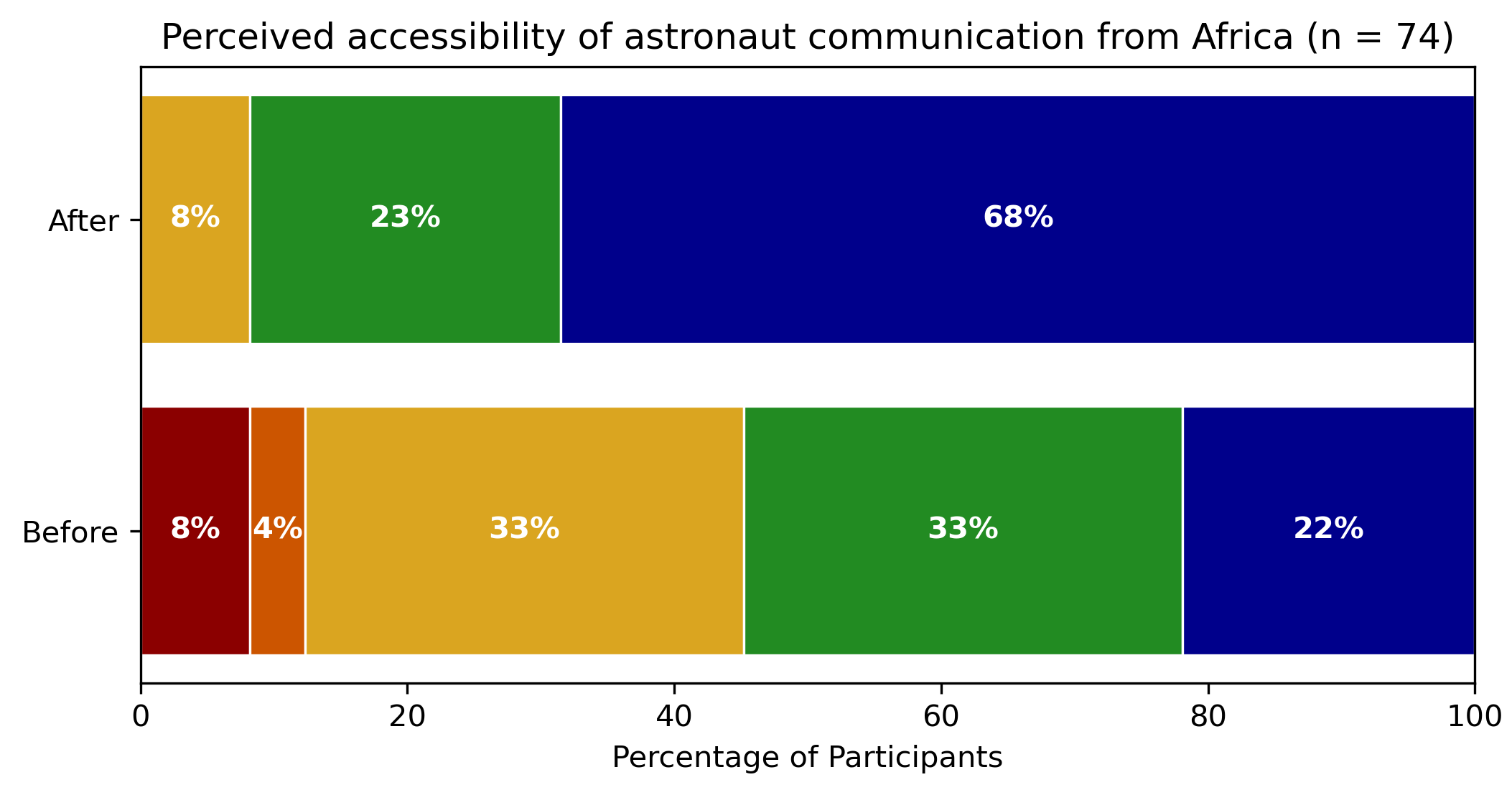}

\includegraphics[width=0.5 \textwidth]{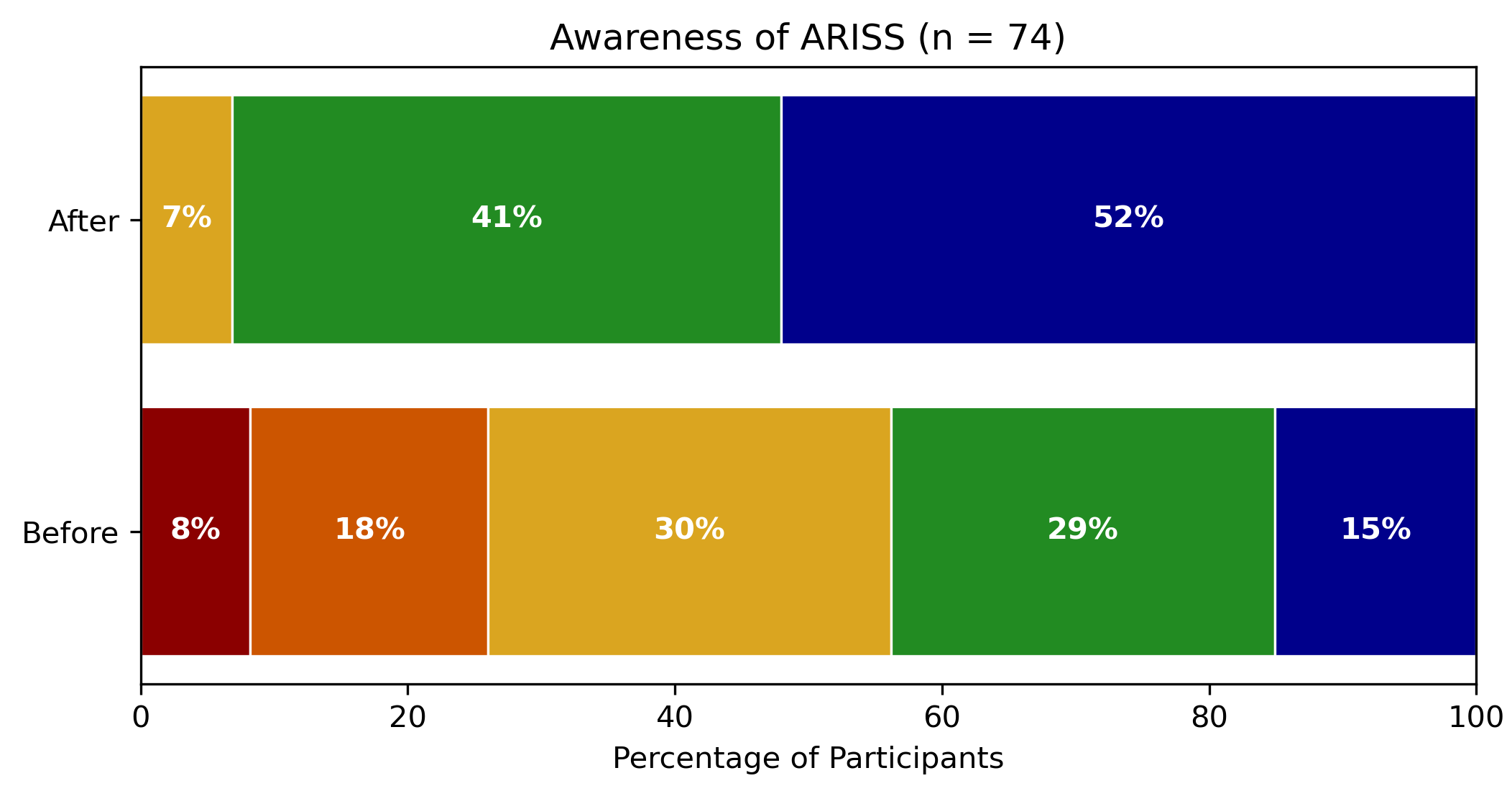}

\includegraphics[width=0.5 \textwidth]{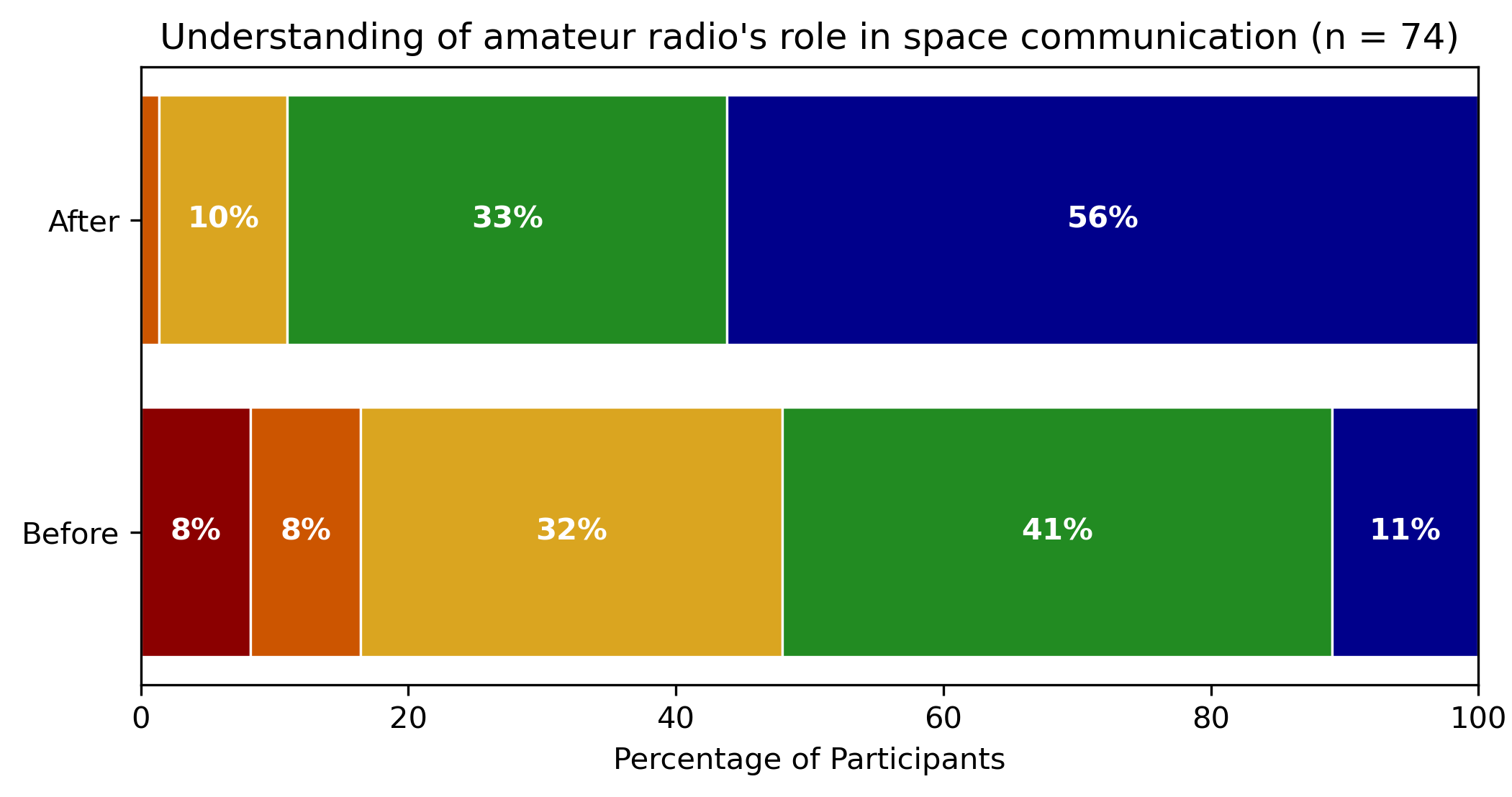}

\caption{Survey: ARISS experiment impact on ISS Awareness, the importance of ham radio and the perceived accessibility of astronaut communication to an African audience.}\label{fig:graph2}
\end{figure}

Feedback surveys were administered to all participants using a five-point Likert scale (1 = "Strongly Disagree" to 5 = "Strongly Agree"). A total of 73 participants completed both the pre- and post-event questionnaires. Before the ARISS contact, participants already demonstrated a high level of interest in astronomy and space science, with most respondents indicating positive attitudes toward space exploration and STEM-related topics. As shown in Figures \ref{fig:graph1},\ref{fig:graph2} responses shifted consistently toward the positive end of the scale following the ARISS contact. Across all measured domains, including interest in space exploration, knowledge of the ISS, awareness of the ARISS program, interest in STEM subjects, perceptions of the accessibility of communicating with astronauts from Africa, curiosity about space-related careers, and understanding of amateur radio's role in space communication, the proportions of participants selecting "Agree" and "Strongly Agree" increased substantially after the event, while "Disagree" and "Strongly Disagree" responses decreased or disappeared almost entirely. These findings, expressed in Figures \ref{fig:graph1},\ref{fig:graph2} indicate that participation in the ARISS contact positively influenced participants' knowledge, attitudes, and enthusiasm towards ISS, space science and STEM education.

Another key observation was the level of excitement and pride expressed by most participants upon receiving their certificates of participation. We list below some of Survey highlight responses:
\begin{itemize}
    \item Vipparthi Keerthi (India): “Participating in the ARISS contact event was
a truly eye-opening experience. . . Hearing the astronaut’s voice live from
orbit helped me realize how accessible space communication has become.
It reminded me that space exploration is not only about technology and
science but also about community, curiosity, and the human desire to
connect—no matter how far apart we are.”
    \item Shiwa Quraishi (Afghanistan): “One of the most interesting parts was the
meeting with an astronaut. It was an incredible and unforgettable memory
for me. At that moment, I was so excited, it felt like I was touching my
dreams, and I said, ‘Oh my God!’ I truly felt happy and more motivated.”
    \item Promise Orolisen (Nigeria): “As space enthusiasts gathered before the
contact time, excitement and anticipation filled the air. The event ignited
curiosity and enthusiasm, inspiring participants to pursue careers
in science. More, it offered a rare chance to learn from someone with
space experience, providing valuable insights into NASA’s communication
processes.”
    \item Veronicah Nyambura Kihagi (Kenya): “On April 18, 2025, I attended the
first-ever radio amateur contact from the ISS to Africa. It was a powerful
reminder that science, technology, and STEM education have no borders.
This first ISS radio contact with Africa is a stepping stone for future
collaborations. It ignites dreams among young people and will forever be
a milestone in my journey of learning and discy.”
    \item Ssebaggala Chrizestom (Uganda): “April 18, 2025, was an exciting and
unforgettable day to listen to an astronaut’s voice directly from aboard
the ISS. I promise to extend the knowledge and facts learned from the ISS
to schools in Uganda and to policy makers to embrace astronomy.”
    \item Amhika Bhugwandass (South Africa): “The experience was truly out of
this world. I never thought anyone outside of space communications could speak with an astronaut on the ISS. It was amazing to have what can only be called a once-in-a-lifetime experience, and I hope to be part of future projects that will deepen my love for the universe.”
    \item Biruk Ayalew Deribe (Ethiopia): “Attending this event provided valuable
insights into the collaborative efforts between African and European space
science initiatives. Participants learned about the critical role of ground
stations, the growing influence of African astronomy, and the importance
of platforms like PACS e-Lab in fostering STEM education across continents.”
\end{itemize}

These emotional responses underscore the profound psychological and inspirational
value of direct connections with space missions and ISS in particular and are 
an indicator of the long-term positive impact on their educational and career aspirations. 

\section{Discussion and Conclusions}

The outlined partnership between PACS e-Lab and ARISS for space education \& outreach met a common educational mission milestone: {\it "To inspire the next generation of explorers"}. 

The successful execution of the first Africa-wide ARISS experiments with PACS e-Lab and partners around the commemorations of the International Day of Human Space Flight (12th April 2025) and during the week of the official inauguration of the African Space Agency, highlights several key achievements, challenges, and lessons that can inform future initiatives of this nature across the continent. Comparing the significance of pre- and post-event interest and excitement levels, there was a noticeable positive trend on space education interest involving the ISS.

Hence, we demonstrated the power of regional collaboration (with existing entities in Africa) and external collaboration (with ARISS and other partners) in mobilizing widespread participation across geographically underrepresented communities with about 339 individuals from 38 countries in Africa and beyond. This illustrates how effective networking, shared goals and digital platforms can overcome infrastructural barriers that traditionally limit STEM outreach and space science education in Africa. Among the many participants, we highlight the participation for the first time of a young woman from Afghanistan, the first woman from this country to interact with an astronaut in orbit. 

Going forward, PACS e-Lab is building on this momentum through a multi-pronged approach: besides developing a campaign for the adoption of ARISS experiments across institutions throughout Africa, it keeps the exposure to projects outlined above, some of which are conducted using PACS e-Lab’s access to robotic telescopes like the 0.4m Las Cumbres Observatory \cite{Brown_2013}, MicroObservatory \cite{Gould2006}, and Slooh \cite{Slooh2023}. These are activities aligned with the vision set by the Science strategy of the African Astronomical Society and the International Astronomical Union \cite{Leeuw_Govender_Takalana_Randriamanakoto_Mamo_2019}.

\begin{acks}
PACS e-Lab are deeply appreciative to NASA and NASA Astronaut Nichole Ayers for her great collaboration.
Our heartfelt thanks are due to the African Astronomical Society, Africa2Moon and Foundation for Space Development Africa, Alliance for Collaboration in the Exploration of Space – ACES Worldwide and HCP Chair at U.Évora. PACS e-Lab acknowledges support from ISC Intelligence in Science. The authors acknowledge space agency sponsors NASA, ESA, CSA, JAXA, and Roscosmos, who have been instrumental in realizing the ARISS educational outreach program. Special recognition goes to the international ARISS volunteer team for their tireless efforts in making ARISS such a successful, affordable space related STEM initiative. 
\section*{Author Contributions} 
\textbf{DB., M.C.M.} Methodology, Writing -- original draft. \textbf{M.C.M.} Conceptualization, Methodology, Data curation, Formal analysis, \textbf{D.K., D.B., M.A., M.C.M.} Funding acquisition, \textbf{C.A., S.D., P.K., M.C.M.} Supervision, Methodology, \textbf{C.S.M., A.A.G., V.B.} Event Participation, Reviewing.
\end{acks}

\begin{dci}
No competing financial interests exist.\\
\end{dci}

\begin{funding}
DB and MA acknowledge support from ENGAGE-SKA Research Infrastructure, ref. POCI-01-0145-FEDER-022217 funded by COMPETE 2020 and FCT – Fundação para a Ciência e a Tecnologia, I.P., Portugal and from the HPC Chair at University of Évora. ARISS support is provided by NASA, the Space Station Explorers consortium and the ISS USA National Laboratory through Space Act Agreement SAA-OZ-21-33848, awarded by NASA Johnson Space Center. \\

\end{funding}
\theendnotes
\bibliographystyle{Sagev}
\bibliography{ARISS_AA_TN}{}

\end{document}